\documentclass[10pt]{article}

\usepackage[letterpaper,margin=0.75in,top=0.7in,bottom=0.75in]{geometry}
\usepackage[T1]{fontenc}
\usepackage[utf8]{inputenc}

\usepackage{lmodern}
\usepackage{amsmath,amssymb,amsfonts,mathtools}
\usepackage{graphicx}
\usepackage{booktabs}
\usepackage{multirow}
\usepackage{xcolor}
\usepackage{microtype}
\usepackage{enumitem}
\usepackage{caption}
\usepackage{subcaption}
\usepackage{url}
\usepackage{xurl}
\usepackage[numbers,sort&compress]{natbib}
\usepackage{tikz}
\usetikzlibrary{positioning,arrows.meta,calc,fit,backgrounds}
\usepackage{titlesec}
\usepackage{fancyvrb}
\usepackage{needspace}
\usepackage{hyperref}

\definecolor{cAgent}{HTML}{FFE7A8} 
\definecolor{cTool}{HTML}{CDE9D6}  
\definecolor{cDisp}{HTML}{E6E2EF}  
\definecolor{cPhys}{HTML}{CFE3F5}  
\definecolor{cNote}{HTML}{E2D4F0}  
\definecolor{cScore}{HTML}{F6CFD4} 
\definecolor{cAux}{HTML}{ECEDEF}   
\definecolor{cEdge}{HTML}{3E4651}  

\definecolor{accentDark}{HTML}{2B3A4A}   
\definecolor{accentRule}{HTML}{5B7FA6}   
\definecolor{accentGray}{HTML}{6B7280}   

\tikzset{
  dgmbox/.style={rounded corners=2.2pt, draw=cEdge, line width=0.5pt,
    align=center, inner sep=4pt, font=\footnotesize},
  dgmagent/.style={dgmbox, fill=cAgent, line width=0.9pt},
  dgmtool/.style={dgmbox, fill=cTool},
  dgmdisp/.style={dgmbox, fill=cDisp},
  dgmphys/.style={dgmbox, fill=cPhys},
  dgmnote/.style={dgmbox, fill=cNote},
  dgmscore/.style={dgmbox, fill=cScore},
  dgmaux/.style={dgmbox, fill=cAux},
  dgmflow/.style={-{Stealth[length=2.2mm,width=1.8mm]}, draw=cEdge, line width=0.7pt},
  dgmflowback/.style={dgmflow, densely dashed},
  dgmlbl/.style={font=\scriptsize\itshape, text=cEdge, inner sep=1.5pt, fill=white,
    fill opacity=0.85, text opacity=1},
}

\titleformat{\section}
  {\needspace{5\baselineskip}\normalfont\sffamily\Large\bfseries\color{accentDark}}
  {\thesection}{0.6em}{}
\titleformat{\subsection}
  {\needspace{4\baselineskip}\normalfont\sffamily\large\bfseries\color{accentDark}}
  {\thesubsection}{0.6em}{}
\titleformat{\paragraph}[runin]
  {\normalfont\sffamily\bfseries\color{accentDark}}
  {}{0em}{}
\titlespacing*{\section}{0pt}{0.9em}{0.45em}
\titlespacing*{\subsection}{0pt}{0.7em}{0.3em}
\titlespacing*{\paragraph}{0pt}{0.65em}{0.4em}

\hypersetup{
  colorlinks=true,
  linkcolor=accentRule,
  citecolor=accentRule,
  urlcolor=accentRule,
  pdftitle={Self-Specializing Vision-Language Transmon Chip Calibration in a Physics-Grounded Environment},
  pdfauthor={Animesh Tripathy, Aswanth Krishnan},
}

\newcommand{\code}[1]{\texttt{\detokenize{#1}}}
\newcommand{\budgetFigWidth}{0.6\linewidth}

\title{\textbf{Self-Specializing Vision-Language Transmon Chip Calibration in a Physics-Grounded Environment}}

\author{Animesh Tripathy \and Aswanth Krishnan \\ QpiAI India Pvt. Ltd.}

\begin{document}
\pagenumbering{arabic}
{\centering
\vspace*{-1.2em}
{\sffamily\bfseries\LARGE\color{accentDark} Self-Specializing Vision-Language Transmon Chip Calibration in a Physics-Grounded Environment\par}
\vspace{1.0em}
{\sffamily\large Animesh Tripathy \quad Aswanth Krishnan\par}
\vspace{0.3em}
{\sffamily\normalsize\color{accentGray} QpiAI India Pvt. Ltd.\par}
\vspace{0.9em}
\par}
\thispagestyle{plain}

\begin{abstract}
Calibrating a superconducting transmon chip is a sequential decision problem under noise, drift, and a finite measurement budget, requiring an expert to choose experiments, read ambiguous plots, judge fit quality, and revise stale beliefs as the chip drifts. We study whether a vision-language agent can close this loop and specialize itself to one physical device without weight updates, via three co-designed artifacts.
The first is \textbf{a physics-grounded simulation environment} for transmon chips: calibration observables ($f_{01}$, $\alpha$, $\chi$, $g$, $\zeta$) derive from circuit-quantized parameters via scqubits, with realistic flux-line distortion, wall-time-scaled drift channels, mid-scan drift, and gate leakage; these are concerns a toy depolarizing simulator would omit, and each tool call advances a modeled measurement clock so drift accrues by wall time rather than by call count.
The second is \textbf{a vision-language calibration agent} that runs the loop end to end, calling measurement tools, reading returned plots, maintaining a structured notebook, and submitting parameters without hidden truth, scored against both hidden parameters and gate fidelities measured on the device.
The third is \textbf{gradient-free online adaptation}: a reflector reads truth-free anomaly signatures from past attempts and grows a small, human-readable device note appended to the agent's prompt, admitted by a paired-snapshot accept gate that isolates strategy improvement from drift.
On a hard-tier chip under budget pressure, six iterations of online adaptation raised the worst-case measured CZ fidelity from $0.678$ to $0.787$ and cut its variance, an effect that reproduces at four-qubit scale; a single accepted note raised CZ fidelity from $0.678$ to $0.913$ on its paired snapshot. A planted-fault study confirms the note is causal, diagnosing a specific hardware fault truth-free, and that its principal value is raising the failure floor and cutting variance.
The environment is not a hardware digital twin, but the agent, scoring, and reward transfer to real hardware via a measurement-backend swap; only the accept gate is a simulation affordance, reducing on hardware to a held-out-slice or repeat-and-average form.
\end{abstract}

\section{Introduction}
\label{sec:intro}

Closed-loop scientific agents are most valuable when measurement is costly, observations are ambiguous, and the right next action depends on the quality of the previous evidence.
Calibrating a superconducting transmon chip has exactly this structure.
The quantities that define a usable processor drift over time: resonator and qubit transition frequencies, pulse amplitudes, anharmonicity, coherence envelopes, leakage suppression, static $ZZ$ coupling, flux response, crosstalk, and two-qubit-gate operating points.
The transmon architecture is well established~\cite{koch2007transmon,blais2021cqed,krantz2019guide} and the experimental toolkit is mature: spectroscopy, Rabi sweeps, Ramsey and echo measurements, DRAG tuning~\cite{motzoi2009drag,gambetta2011drag}, randomized benchmarking~\cite{magesan2011rb}, quantum volume~\cite{cross2019qv}, cross-entropy benchmarking~\cite{boixo2018xeb,arute2019supremacy}, and throughput metrics such as CLOPS~\cite{wack2021clops}.
The hard part is not any single measurement. It is sequencing measurements under evidence quality, drift, fit failures, crosstalk, and a finite budget, and then re-running the loop as the chip ages away from its last calibration.
Automated calibration tools and optimal-control routines~\cite{wittler2021toolset,majumder2020spectator} have targeted individual calibration steps, but orchestrating them into an open-ended agent loop that adapts to a specific device as it drifts remains an open problem.
Concurrent work shows that a language agent can autonomously bring up a real 112-qubit processor by orchestrating a curated library of expert-authored calibration ``skills''~\cite{xu2026vibe}, reflecting the timeliness of agentic calibration.
Our emphasis is complementary: a \emph{vision-language} policy that reads rendered measurement plots as primary evidence, is scored against hidden ground truth, and is specialized to one persistent drifting device by a gradient-free online loop rather than by a fixed, hand-authored skill library.

This paper asks whether a tool-using vision-language model can serve as the calibration \emph{policy} for a whole transmon chip, and, crucially, whether that policy can be \emph{specialized to one specific physical device without any weight updates}, by growing a small, interpretable, natural-language note that the agent reads.
We pursue this through three tightly co-designed artifacts: a physics-grounded environment, a vision-language agent that runs the loop end to end, and a gradient-free loop that adapts the agent online to a single drifting chip.

We make the policy vision-language so that the agent receives the same evidence a human calibration engineer reads: each tool response returns a rendered figure (a spectroscopy peak, a Rabi oscillation, a chevron heatmap) alongside a compact numeric summary.
In principle the figure carries information a bare scalar can hide, such as whether a fit locked onto a real transition or a TLS artifact, whether a window bracketed a full oscillation, or whether a scan's fringes slanted under mid-scan drift.
This information can be decisive, but it is masked whenever the tools also return scalar guards (the fit-quality $r^2$, recommended values, and one-line evidence summaries) that mirror the figure into text: simply withholding the plots then costs nothing, because the same content survives in the scalars.
Once those redundant text channels are closed, the plots carry real weight: reading the rendered figure alone recovers $+0.16$ parameter coverage over a blind agent and eliminates the catastrophic zero-coverage failures a blind agent suffers (Section~\ref{sec:exp-ablation}).
The multimodal channel is therefore both \emph{necessary} for parameter recovery when evidence is not pre-digested into text and \emph{auditable}, letting an operator see why each value was committed.
Concurrent work isolates this perception skill directly: \citet{cao2026qcaleval} benchmark whether vision-language models can read quantum calibration plots in a static question-answering setting, finding that even frontier models leave substantial headroom; our contribution is complementary, embedding plot-reading inside a closed calibration loop that acts on a device and is scored against hidden ground truth.
This places the policy near the ``reasoning-plus-acting'' line of LLM-agent work~\cite{yao2023react,schick2023toolformer,patil2024gorilla,qin2024toollm} and the use of agents as autonomous scientific experimenters~\cite{bran2024chemcrow,boiko2023autonomous}, but in a multimodal control domain where every action has a physics interpretation and the environment can score against hidden ground truth.

Environment fidelity earns its place as a co-equal contribution. A learned calibration policy is only useful if its strategies transfer to real hardware, which makes the noise and physics model a core research deliverable rather than development scaffolding.
Toy depolarizing noise produces an agent that never has to confront the failure modes that dominate a real calibration shift: flux pulses that ring and settle slowly, fits that degrade because the qubit drifted mid-scan, leakage that caps achievable gate fidelity, and parameters that go stale at different wall-time rates.
Our environment makes each of these present, physically grounded, and independently ablatable, so that conclusions about the agent's policy are conclusions about behavior that could transfer to a deployed chip.

Three deployment constraints push the adaptation onto a gradient-free, online, and interpretable form. On real hardware there is no ground truth to backpropagate through, no practical way to fine-tune a frontier model per device, and a strong operational requirement that the control policy be auditable.
We therefore represent the agent's device-specific strategy as a small natural-language note, separated from the frozen base prompt, that the agent reads but that an outer loop edits.
Over repeated calibration attempts on the \emph{same} persistent, drifting chip, a reflector inspects truth-free anomaly signatures from recent attempts and proposes edits to this note, deliberately \emph{overfitting} to the one device in front of it (for example, ``this chip's flux line is unusually fast, so use a denser cryoscope grid'').
A paired-snapshot accept gate then admits an edit only if it improves a truth-free, measured objective on the \emph{same} frozen device state, which separates genuine strategy improvement from the device simply having drifted.
The result is a learning curve over calibration episodes whose ``policy delta'' is a human-readable text block, an unusually interpretable form of online adaptation and the test-time, warm-up half of a deployment story.

\paragraph{Contributions.}
\begin{enumerate}[leftmargin=*,nosep]
  \item \textbf{A physics-grounded calibration environment} (Section~\ref{sec:environment}) for superconducting transmon chips: circuit-quantization device truth, realistic flux-line distortion with FIR predistortion, six wall-time-scaled drift channels, mid-scan drift, and CZ leakage, each independently ablatable and motivated by hardware transfer.
  \item \textbf{A vision-language calibration agent} (Section~\ref{sec:agent}) that calibrates a whole chip by reading measurement plots, maintaining a structured notebook, committing parameters under evidence and fit-quality gates, and scoring itself on parameter recovery \emph{and} on gate fidelities measured directly on the device.
  \item \textbf{Gradient-free online specialization} (Section~\ref{sec:optimization}): a sequential loop that grows an interpretable, truth-free device note over repeated calibration attempts on one persistent drifting chip, with a paired-snapshot accept gate, without touching model weights.
  \item \textbf{Empirical results} (Section~\ref{sec:experiments}): a budget-pressure sweep shows the device note raises the worst-case measured CZ fidelity and cuts its variance, validated causally by planting concrete hardware faults behind an ablation flag; an environment-fidelity ablation shows a toy simulator is blind to gaps in measured two-qubit fidelity and predistortion residual that the physics-grounded environment exposes; and a scalar-deprived test shows the rendered plot is essential for parameter recovery.
\end{enumerate}

\section{A Physics-Grounded Calibration Environment}
\label{sec:environment}

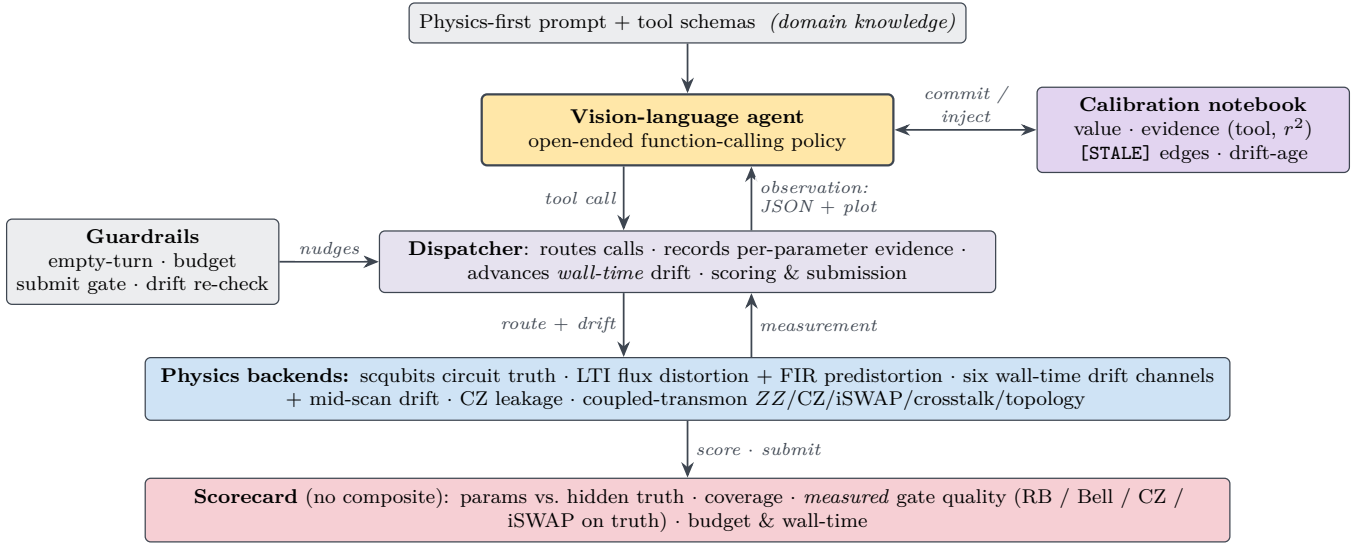
\begin{figure*}[t]
  \centering
  \resizebox{\linewidth}{!}{%
  \begin{tikzpicture}[node distance=4mm]
    \node[dgmaux, minimum width=52mm] (know)
      {Physics-first prompt $+$ tool schemas \;\textit{(domain knowledge)}};
    \node[dgmagent, below=7mm of know, minimum width=58mm, minimum height=10mm] (agent)
      {\textbf{Vision-language agent}\\open-ended function-calling policy};
    \node[dgmdisp, below=9mm of agent, text width=84mm] (disp)
      {\centering\textbf{Dispatcher}: routes calls $\cdot$ records per-parameter evidence
       $\cdot$ advances \emph{wall-time} drift $\cdot$ scoring \& submission\par};
    \node[dgmphys, below=9mm of disp, text width=150mm] (phys)
      {\centering\textbf{Physics backends:} scqubits circuit truth $\cdot$ LTI flux distortion $+$ FIR predistortion
       $\cdot$ six wall-time drift channels $+$ mid-scan drift $\cdot$ CZ leakage
       $\cdot$ coupled-transmon $ZZ$/CZ/iSWAP/crosstalk/topology\par};
    \node[dgmnote, right=20mm of agent.east, anchor=west, minimum width=44mm] (nb)
      {\textbf{Calibration notebook}\\value $\cdot$ evidence (tool, $r^2$)\\\code{[STALE]} edges $\cdot$ drift-age};
    \node[dgmaux, left=14mm of disp.west, anchor=east, minimum width=34mm] (guard)
      {\textbf{Guardrails}\\empty-turn $\cdot$ budget\\submit gate $\cdot$ drift re-check};
    \node[dgmscore, below=8mm of phys, text width=150mm] (score)
      {\centering\textbf{Scorecard} (no composite): params vs.\ hidden truth $\cdot$ coverage
       $\cdot$ \emph{measured} gate quality (RB / Bell / CZ / iSWAP on truth) $\cdot$ budget \& wall-time\par};

    \draw[dgmflow] (know) -- (agent);
    \draw[dgmflow] ([xshift=-9mm]agent.south) -- node[dgmlbl,left=0.5mm]{tool call} ([xshift=-9mm]disp.north);
    \draw[dgmflow] ([xshift=9mm]disp.north) -- node[dgmlbl,right=0.5mm]{\shortstack[l]{observation:\\JSON $+$ plot}} ([xshift=9mm]agent.south);
    \draw[dgmflow] ([xshift=-9mm]disp.south) -- node[dgmlbl,left=0.5mm]{route $+$ drift} ([xshift=-9mm]phys.north);
    \draw[dgmflow] ([xshift=9mm]phys.north) -- node[dgmlbl,right=0.5mm]{measurement} ([xshift=9mm]disp.south);
    \draw[dgmflow,{Stealth[length=2.2mm]}-{Stealth[length=2.2mm]}]
      (agent.east) -- node[dgmlbl,above]{\shortstack{commit /\\ inject}} (nb.west);
    \draw[dgmflow] (guard.east) -- node[dgmlbl,above]{nudges} (disp.west);
    \draw[dgmflow] (phys.south) -- node[dgmlbl,right=0.5mm]{score $\cdot$ submit} (score.north);
  \end{tikzpicture}}
  \caption{Closed-loop calibration environment. Domain knowledge lives in the physics-first prompt and the measurement tools; each turn the vision-language agent calls a tool and receives a multimodal observation (numeric summary $+$ rendered plot). The dispatcher routes calls to the physics backends, records per-parameter evidence in the notebook, and advances device drift by modeled \emph{wall time}; only narrow guardrails are injected, never a scripted policy. The structured scorecard carries no composite and reports gate fidelities \emph{measured on the truth device with calibrated parameters}. The same tool interface is designed to back a real-hardware shim.}
  \label{fig:closed-loop}
\end{figure*}

The central design commitment of this work is that the simulator, noise model, and benchmark suite are co-equal research deliverables with the agent.
A calibration strategy is only worth learning if it transfers, and a strategy learned against toy noise does not have to solve the problems that dominate a real calibration shift.
This section describes the simulation environment as a layered, physically grounded, and independently ablatable stack for whole superconducting transmon chips.
Every realism feature below carries an \emph{ablation flag} that recovers the simpler behavior bit-for-bit, which is what lets us attribute agent behavior to a specific physical effect.

\subsection{Architecture and the Prompt/Runtime Split}

The environment has four layers (Fig.~\ref{fig:closed-loop}).
The agent loop owns the vision-language function-calling conversation: each turn may call one or more tools, and each tool response returns compact JSON plus a rendered PNG plot as multimodal function-response data.
The dispatcher routes calls to physics simulators, records per-parameter evidence, advances drift by modeled hardware wall time, and handles scoring and submission.
The notebook maintains a parameter graph in which each parameter records its value, the evidence tool that set it, a one-line evidence summary, the fit quality ($r^2$) that produced it, its set-step, a revision count, and a stale flag propagated along hard-coded dependency edges.
The physics backends provide analytical sweeps for high-volume scans, QuTiP Lindblad simulation~\cite{johansson2012qutip,johansson2013qutip2} for gate-level dynamics, scqubits-derived circuit truth, and coupled-transmon simulators for flux response, $ZZ$, CZ, iSWAP, crosstalk, leakage, and chip topology.

Domain knowledge is deliberately placed in physics-first prompts and measurement tools rather than a scripted recipe.
The runtime supplies only narrow operational support: recover from empty turns, enforce budget and submission, avoid pathological repeated single-qubit re-scans, and prompt targeted drift re-checks when aged parameters may matter.
This makes the environment a testbed for experimental \emph{policy}: when should an agent trust a fit, spend another measurement, revise a stale belief, or stop?

\subsection{Circuit-Quantization Device Truth}
\label{sec:truth}

Sampling calibration-level observables ($f_{01}, \alpha, \chi, g, \zeta$) as independent uniforms produces physically inconsistent devices and lets an agent ignore relationships that real calibration teams exploit.
The environment instead derives truth from circuit quantization (scqubits~\cite{groszkowski2021scqubits}).
Each transmon is sampled at the circuit level, as $(E_C, E_J, d, n_g)$ together with a capacitive coupling $g_{qq}$ for neighbors, and the observables follow from diagonalization:
\[
  f_{01} \approx \sqrt{8 E_C E_J} - E_C, \qquad \alpha \approx -E_C,
\]
with $g$ from charge matrix elements and the static $ZZ$ rate $\zeta$ from \emph{full} two-transmon diagonalization rather than the leading-order dispersive expression~\cite{blais2021cqed}
\begin{equation}
  \zeta \approx 2g^2\!\left[\frac{\alpha_1}{\Delta(\Delta+\alpha_1)} - \frac{\alpha_0}{\Delta(\Delta-\alpha_0)}\right],
  \label{eq:zz}
\end{equation}
which breaks down precisely at the $|11\rangle\!\leftrightarrow\!|02\rangle$ avoided crossing where the CZ gate operates.
The agent never sees $(E_C, E_J)$; it sees only the now self-consistent observables.
Tunable transmons expose an exact $f_{01}(\Phi)$ flux curve (including junction asymmetry $d$) via eigenvalue solves, which the flux-distortion layer (Section~\ref{sec:flux}) convolves against.
Coherence times ($T_1, T_2, T_2^*$), TLS defects, and readout asymmetry remain empirical overlays, since they are environment-dominated rather than circuit-derived.
Difficulty tiers (easy/medium/hard) are specified as circuit-parameter and noise ranges, and the chip sampler enforces collision avoidance ($\ge 100$~MHz between neighbors) and dispersive detuning (350--650~MHz per edge).

\subsection{Realistic Flux-Line Distortion and Predistortion}
\label{sec:flux}

On real hardware, every commanded flux pulse passes through a transfer function (line inductance, DAC bandwidth, bias-tee high-pass, filter settling) before reaching the coupler; cryoscope~\cite{rol2020cryoscope} measures this step response and predistortion inverts it.
This is a critical realism feature: without it, predistortion parameters have no downstream effect and the entire flux-dependent gate-calibration stage is cosmetic.

The flux line is modeled as a composable, physically typed distortion kernel (a dominant fast pole, an optional slow cryo-tail pole, a bias-tee high-pass droop, and, on the hard tier, an AWG LC ring), applied by \emph{true LTI convolution}, so that both the rising and falling pulse edges and the post-pulse settling tail are exposed.
The magnitudes are set to a current-hardware regime in which the bulk of the distortion occurs in the first $\sim$20~ns, so the dominant pole is fast ($\tau \in [10, 25]$~ns) and the gate sees a mostly-settled flux with a small tail, as on a real device.\footnote{Magnitudes are calibrated to~\cite{venkateswaran2026dpd}, which characterizes flux-control distortion on a current superconducting processor and finds the bulk of distortion within the first $\sim$20~ns.}
The agent calibrates against this by fitting cryoscope data with either a single-pole model or a \emph{multi-tap FIR filter} (Tikhonov deconvolution), the latter capturing ringing a single pole cannot.
Predistortion is scored on the achieved residual RMS between commanded and realized flux, the quantity that actually limits gate fidelity, rather than on a scalar time constant.
The ablation flag (identity kernel) recovers ideal square pulses bit-for-bit.

\subsection{Wall-Time-Scaled and Mid-Scan Drift}
\label{sec:drift}

Real calibrations are paced by a wall-clock: a randomized-benchmarking sweep takes minutes while an AllXY takes seconds, and a parameter's validity is tied to elapsed time, not to how many tool calls have occurred.
The environment gives every charged measurement a modeled hardware duration (e.g.\ $\sim$21~s resonator spectroscopy, $\sim$8~min CZ chevron) and accumulates a cumulative wall time.
Drift then scales with wall time: a tool taking $w$ seconds advances each channel by $w/T_{\text{ref}}$ (deterministic steps) and $\sqrt{w/T_{\text{ref}}}$ (stochastic walk, Brownian rescaling), while zero-cost tools (parameter setting, fitting, reasoning, submission) do not drift the device.
Six drift channels, each independently ablatable, layer on top of the base frequency drift: drive-amplitude drift (the most-recalibrated quantity on real hardware), asymmetric readout-discriminator walk, TLS spectral diffusion ($T_1$ jitter)~\cite{klimov2018fluctuations}, slow $1/f$ flux offset~\cite{ithier2005decoherence}, coupler/$ZZ$ bias-point walk, and a shared local-oscillator source that injects $>0.5$ Pearson-correlated drift across qubits sharing a line.
The slow-flux channel is \emph{operating-point dependent}: rather than shifting the idle frequency by a flat offset, the flux excursion is routed through the device's own $f_{01}(\Phi)$ curve at the parked bias, so a tunable transmon held near its sweet spot inherits the first-order flux insensitivity ($\partial f_{01}/\partial\Phi\to 0$) that makes the sweet spot the chosen idle point. The resulting idle-frequency drift is suppressed by $\sim$5$\times$ relative to the slope away from the sweet spot, and vanishes for fixed-frequency qubits, while the TLS, charge, and LO sources (which shift $f_{01}$ directly) remain unsuppressed.

A separate \emph{mid-scan} drift mechanism advances the device \emph{within} a single 2D sweep, one step per column of a chevron, so the avoided-crossing target shifts horizontally and the fringes slant and smear, exactly as a frequency drift mid-acquisition does on hardware.
This degrades the 2D chevron fit $r^2$ on noisy chips and forces the agent to accept fallbacks rather than trusting a clean fit.

At the chip level, drift is \emph{persistent}: while the agent calibrates one qubit or edge, its neighbors age by the full wall time spent, and a shared-LO registry survives across the chip's calibration sub-tasks, so the chip is never re-frozen to a pristine state.
This persistent, drifting chip is precisely what the online adaptation loop (Section~\ref{sec:adapt}) specializes to.

\subsection{Gate Leakage and Measured Gate Quality}
\label{sec:leakage}

CZ chevrons carry leakage into non-computational levels near the avoided crossing.
The environment layers two physical effects on the two-state envelope: perturbative coherent leakage to $|20\rangle$ via off-resonant single-excitation coupling, and a non-adiabatic leakage floor proportional to pulse risetime.
These cap achievable CZ fidelity on the hard tier (peak $P_{11}$ ceiling $<0.985$, versus $>0.99$ on easy), reproducing a real ceiling rather than allowing arbitrarily good gates.

Calibration quality is measured, not assumed.
A dedicated scoring block runs randomized benchmarking, Bell tomography, and CZ/iSWAP characterization \emph{on the truth device with the agent's calibrated parameters installed}.
This is the closest in-environment analog of a real hardware figure of merit, and because it is a measurement rather than a comparison to a hidden number, it is the basis of the truth-free objective used by online adaptation (Section~\ref{sec:adapt}).
The full set of hardware-transfer additions is summarized in Table~\ref{tab:realism}, and Appendix~\ref{sec:appendix-plots} collects representative tool-call plots (resonator and qubit spectroscopy, Hahn echo, cryoscope, $ZZ$ Ramsey, CZ and iSWAP chevrons, Bell-state tomography, and a pulse-level flux-predistortion trace) from hard-tier calibration runs to make the noise floor, TLS structure, and fit-quality judgments described above concrete.

\begin{table}[t]
\centering
\small
\caption{Hardware-transfer realism features, each independently ablatable to recover the simpler behavior. The ablation flags are what let agent behavior be attributed to a specific physical effect.}
\label{tab:realism}
\begin{tabular}{@{}p{0.30\columnwidth}p{0.62\columnwidth}@{}}
\toprule
Feature & What it adds \\
\midrule
Circuit truth & scqubits $(E_C,E_J,d)\!\to\!(f_{01},\alpha,\chi,g,\zeta)$; exact $f_{01}(\Phi)$; full-diagonalization $ZZ$ \\
Flux distortion & LTI convolution of typed kernel; FIR predistortion; residual-RMS scoring \\
Wall-time drift & per-tool durations; drift $\propto$ wall time; persistent chip state \\
Drift channels & 6 ablatable: drive, readout, TLS, slow flux, coupler, shared-LO \\
Mid-scan drift & intra-sweep aging; slanted/smeared chevrons; lower fit $r^2$ \\
CZ leakage & coherent + non-adiabatic; measured-fidelity ceiling \\
Gate quality & RB / Bell / CZ / iSWAP measured on truth with calibrated params \\
\bottomrule
\end{tabular}
\end{table}

\section{The Vision-Language Calibration Agent}
\label{sec:agent}

The agent is a multimodal, tool-using model that calibrates a whole transmon chip end to end without ever seeing hidden truth.
It receives tool schemas, a physics-grounded system prompt, the calibration notebook, and multimodal tool responses, each a numeric summary together with a rendered measurement plot, and from these it must choose experiments, interpret noisy or ambiguous figures, commit calibrated parameters, and submit.

\subsection{Tools and Evidence-Aware Commits}

The tool set mirrors the measurement primitives of real control electronics, so that swapping the simulator backend for hardware is a thin shim rather than a rewrite.
Qubit-characterization tools include resonator and qubit spectroscopy, Rabi and fine-amplitude sweeps, DRAG scans, AllXY, $T_1$, Ramsey, echo, state and process tomography, and randomized benchmarking.
Two-qubit-gate tools add flux sweeps, $ZZ$ Ramsey, CZ and iSWAP chevrons, cryoscope and predistortion fitting, joint-flux-sweetspot crosstalk, Bell tomography, and microwave crosstalk.
A chip is calibrated by composing these over its qubits and edges (Section~\ref{sec:orchestration}).

Each measurement and fit tool returns a fit-quality record $\{r^2, \text{residual\_rms}, n_\text{points}, \text{window\_used}\}$.
The prompt gates commitment on $r^2$: trust $r^2 \ge 0.9$, verify $0.5 \le r^2 < 0.9$, and avoid committing $r^2 < 0.5$ unless a justified fallback is required.
Several gate-calibration tools return guarded \emph{derived} estimates (e.g.\ \code{recommended_cz_duration_us} from a 2D chevron fit, $g$ from a $ZZ$ measurement): when a derived value falls outside a physical range the tool nulls it and emits a warning, and the prompt instructs the agent to fall back to a documented prior.
This is a deliberate robustness mechanism, since chained estimates without guards propagate one bad fit into several wrong parameters.

\subsection{Reading the Plots}

Because the agent is vision-language, the rendered figure is presented as primary evidence rather than decoration.
A spectroscopy plot reveals whether the peak the fitter locked onto is the real transition or a TLS artifact (Appendix Fig.~\ref{fig:app-qubitspec}); a Rabi plot shows whether the amplitude window bracketed a full oscillation; a chevron heatmap shows whether the fringes have slanted under mid-scan drift (Section~\ref{sec:drift}) and the fit should be distrusted.
Appendix~\ref{sec:appendix-plots} collects such plots, and Appendix~\ref{sec:app-traces} shows the agent acting on them.
The tool responses are therefore multimodal, and the agent is asked to reconcile the figure with the reported fit quality before committing a value.
Because our tools also mirror these disambiguations into scalar guards (fit-quality $r^2$, recommended values, evidence summaries), simply withholding the plots does not degrade calibration; but when those redundant text channels are closed, the plots carry real weight, recovering substantial parameter coverage over a blind agent (Section~\ref{sec:exp-ablation}). The figure is thus both evidence the agent acts on and an auditable record of why each value was committed.

\subsection{Calibration Notebook and History Compression}

The notebook is the authoritative, compact state the agent reasons over.
Beyond the scalar parameter values, it records per-parameter evidence (tool, summary, $r^2$, set-step), hard-coded dependency edges that mark descendants \code{[STALE]} when a parameter is re-set, a \code{revise} operation for rolling back a committed value, and wall-time drift-age tags (\code{age:N}, \code{drift?}) that flag values old enough to warrant re-verification.

Because plot-heavy transcripts exhaust context, the agent compresses history: after each successful parameter commit, recent measurement turns are rewritten in place to one-line evidence summaries and their PNGs are dropped, while the model's own reasoning turns are preserved.
The prompt explicitly instructs the agent to trust the notebook over scrolling history.
Appendix~\ref{sec:app-notebook} reproduces a rendered notebook captured mid-calibration, with its evidence records, drift-age tags, and stale-marked parameters, alongside an example of a compressed measurement turn.

\subsection{Reasoning, Guardrails, and Chip Orchestration}
\label{sec:orchestration}

The agent emits a zero-cost reasoning log (``I know / I need / Next'') as the first action each turn, which makes the policy inspectable without charging the budget; Appendix~\ref{sec:app-traces} reproduces two such traces, showing the agent reject a possible TLS peak, decline a borderline fit and re-measure, and fall back to a physical prior when a fit fails its quality gate.
The runtime injects minimal guardrails: an empty-turn poke, budget-pressure reminders, a hard submit gate when few calls remain, an auto-submit fallback, a redirect away from pathological repeated single-qubit re-scans, and a single drift re-check once the coupling/$ZZ$/CZ parameters are set.
To calibrate a full chip, an orchestrator runs an isolated, context-clean calibration sub-task per qubit and per edge, accumulating results into a calibration database; sub-tasks never share transcript state, so each is a clean unit for both scoring and note injection (Section~\ref{sec:optimization}).

\subsection{Scoring: Parameter Recovery and Downstream Utility}

Each calibration produces a structured scorecard rather than a single scalar (Table~\ref{tab:score}).
It reports per-parameter hidden-truth errors and within-tolerance status, coverage, \emph{measured} gate quality (RB / Bell / CZ / iSWAP run on the truth device with calibrated parameters), budget and wall time, and submission validity.
There is deliberately no composite field.
When a single scalar is needed for retry triage or for the adaptation accept gate, the convention $0.5\cdot\text{coverage} + 0.5\cdot\text{measured\_fidelity}$ is computed \emph{outside} the scorecard.
Reporting both is intentional: the agent is judged both by whether it recovered the hidden physics and by whether its calibration actually supports downstream circuit execution.

\begin{table}[t]
\centering
\small
\caption{Scorecard fields. The gate-quality block is measured on the truth device with calibrated parameters and is the basis of the truth-free online objective (Section~\ref{sec:adapt}).}
\label{tab:score}
\begin{tabular}{@{}p{0.30\columnwidth}p{0.62\columnwidth}@{}}
\toprule
Field & Meaning \\
\midrule
Params & Per-parameter value, truth, tolerance, within-tol \\
Coverage & Required / set / missing / within-tol counts \\
Gate quality & RB / Bell / CZ / iSWAP measured on truth device \\
Budget & Tool calls, revisions, modeled wall time \\
Submission & Submitted, valid, failure reason \\
\bottomrule
\end{tabular}
\end{table}

\section{Gradient-Free Online Specialization}
\label{sec:optimization}
\label{sec:adapt}

\begin{figure*}[t]
  \centering
  \resizebox{\linewidth}{!}{%
  \begin{tikzpicture}[node distance=10mm]
    \node[dgmnote, text width=30mm] (note)
      {\textbf{Device note}\\additive prompt slot;\\starts empty; per-qubit\\/ per-edge strategy};
    \node[dgmagent, text width=30mm, right=11mm of note] (agent)
      {\textbf{Frozen agent}\\calibrates the chip\\(note appended to\\every sub-task)};
    \node[dgmtool, text width=30mm, right=11mm of agent] (sig)
      {\textbf{Truth-free signatures}\\fit-quality patterns\\$\cdot$ stale/missing params\\$\cdot$ measured-fidelity dips};
    \node[dgmnote, text width=30mm, right=11mm of sig] (refl)
      {\textbf{Reflector}\\(stronger model)\\proposes a \emph{challenger}\\note edit};
    \node[dgmscore, text width=36mm, right=11mm of refl] (gate)
      {\textbf{Paired-snapshot accept gate}\\freeze snapshot;\\incumbent vs.\ challenger;\\accept iff challenger wins};
    \node[dgmphys, text width=152mm, below=15mm of sig] (dev)
      {\textbf{Persistent drifting chip}: one device, never re-drawn; frozen for the paired evaluation, then aged by inter-iteration wall-time drift each loop (Section~\ref{sec:drift}). Objective is \emph{truth-free}: measured gate fidelity (RB, Bell/CZ) $+$ fraction validly submitted $+$ budget.};

    \draw[dgmflow] (note) -- (agent);
    \draw[dgmflow] (agent) -- (sig);
    \draw[dgmflow] (sig) -- (refl);
    \draw[dgmflow] (refl) -- (gate);
    \draw[dgmflowback] (gate.north) -- ++(0,6mm)
      -| node[dgmlbl, pos=0.25, above]{accept: keep edit \quad / \quad reject: roll back to empty} (note.north);
    \draw[dgmflow,{Stealth[length=2mm]}-{Stealth[length=2mm]}]
      (agent.south) -- node[dgmlbl, left=0.5mm]{run on device} (agent.south |- dev.north);
    \draw[dgmflow] (gate.south) |- node[dgmlbl, pos=0.25, right=0.5mm]{freeze / advance} (dev.east);
  \end{tikzpicture}}
  \caption{Gradient-free online specialization to one drifting chip. The base prompt and agent loop stay frozen; the only mutation surface is a human-readable \emph{device note}. Each iteration, the agent calibrates the chip, truth-free anomaly signatures are distilled from the attempt, and a stronger reflector proposes a challenger note. A paired-snapshot accept gate freezes the device, runs the incumbent and challenger notes on the \emph{identical} drifted snapshot, and admits the edit only if it wins the truth-free objective, isolating note improvement from device drift. The device then ages and the loop repeats.}
  \label{fig:adapt-loop}
\end{figure*}
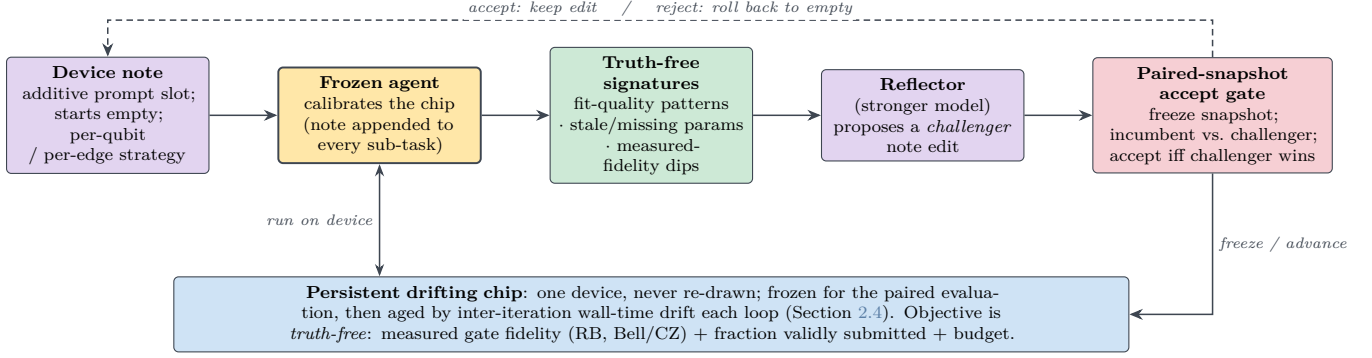

Having a capable agent and a faithful environment raises the question this paper centers on: \emph{can the agent be specialized to one specific physical chip without touching its weights?}
Three constraints from the deployment target make weight training the wrong tool.
First, on real hardware there is no hidden ground truth to backpropagate through.
Second, per-device fine-tuning of a frontier model is impractical and would not be auditable.
Third, the calibration sample budget on a real device is tiny: each ``episode'' is a full calibration shift, so any adaptation must be sample-efficient.

Our answer is to make the agent's device-specific strategy an explicit, isolated, human-readable object, a \emph{device note}, and to grow it with reflective, gradient-free search against the one chip in front of us (Fig.~\ref{fig:adapt-loop}).
The note is the single mutation surface; the base prompt and the agent loop stay frozen.
Unlike a cross-device generalization objective, the goal here is deliberate \emph{specialization}: we overfit the note to a single persistent device on purpose, and rely on a paired accept gate (Section~\ref{sec:gate}) to keep that overfitting in check.

\subsection{The Device Note as an Isolated Mutation Surface}
\label{sec:note}

The one seam into the agent is an additive prompt slot: a string appended to every calibration sub-task prompt under a labeled header.
The empty string reproduces the base agent bit-for-bit; a non-empty note is read by the agent as additional, device-specific guidance.
The note starts empty and is grown by the loop into a short, structured, natural-language document with per-qubit and per-edge sections.
It encodes only \emph{strategy and observed device facts}, such as measurement order, fit-quality discipline, fallback values, and inferred persistent defects, and never hidden answers (the agent never sees ground truth, so this is structurally enforced).
This separation gives three properties at once: the search space is small and the diff is human-auditable; rollback is trivial (empty note $\Rightarrow$ validated base behavior); and ``the agent we adapt is the agent we deploy,'' since there is no forked policy.

\subsection{A Persistent, Drifting Device}

The chip is sampled once and then \emph{mutated in place}: between calibration attempts it ages by a configured wall time using the same wall-time drift used inside episodes (Section~\ref{sec:drift}).
The agent therefore adapts to a moving target, which is the ``evolves with the setup'' requirement of a deployed device.
A fresh device is never re-drawn; all adaptation happens against the history of one chip.

\subsection{Truth-Free Reflective Reward}

The objective must be measurable on the device itself, because a real chip provides no ground truth.
We use only quantities the scorecard measures \emph{on the device}: the gate-quality fidelities (RB, Bell/CZ), the fraction of sub-tasks that validly submitted, and the budget consumed.
Coverage-against-truth is deliberately dropped.
The same reward code path is faithful in both worlds: in simulation we have truth but never read it, and on hardware it is simply absent, so the optimizer cannot accidentally depend on something a real device cannot provide.
A regression test pins this guarantee: poisoning the coverage and truth fields leaves the reward vector bit-identical.

Each iteration, a \emph{reflector}, a stronger model than the agent, reads truth-free anomaly signatures distilled from the most recent calibration attempts (per-qubit and per-edge fit-quality patterns, missing or stale parameters, measured-fidelity dips) over a sliding window, and proposes an edit to the device note in the spirit of textual-feedback prompt optimization~\cite{shinn2023reflexion,yuksekgonul2025textgrad}.
Unlike a generalization-oriented reflector that must keep edits device-agnostic, this one is \emph{encouraged} to write device-specific, per-qubit/per-edge facts inferred from observable anomalies (e.g.\ ``edge (0,1) iSWAP chevron fits are chronically poor, $r^2\approx0.2$; do not over-trust the fitted duration''). It is overfitting one device on purpose; the accept gate is the robustness check.

\subsection{Paired-Snapshot Accept Gate}
\label{sec:gate}

The key mechanism is how an edit is admitted.
Each iteration, after the reflector proposes a new note, the device is \emph{frozen} at its current drifted snapshot, and the agent is run twice on that \emph{same} snapshot: once with the incumbent note and once with the challenger note.
The challenger is accepted only if it beats the incumbent on the truth-free objective (paired across multiple internal seeds, or across a held-out chip slice when the chip is large enough); otherwise the note is rolled back and the rejection logged.
The device only then advances by the inter-iteration drift.

This paired same-snapshot evaluation is what isolates \emph{note improvement} from \emph{device drift}.
Without it, a reward change between iterations could simply be the device having moved rather than the strategy getting better.
Anchoring on unfakeable truth-device fidelities, plus the submission and budget terms, also defends against degenerate ``submit nothing fast'' candidates, and rolling back to an empty note bounds catastrophic forgetting since the base agent is never edited.

We are explicit about which part of this is a simulation affordance.
Freezing the device and replaying the \emph{identical} drifted snapshot under both notes is what gives a clean \emph{causal} attribution of a reward change to the note alone, a controlled-experiment luxury the simulator provides and real hardware cannot, since a physical chip cannot be paused and its drift cannot be rewound.
On hardware the same-snapshot, multi-seed gate degrades to \emph{repeat-and-average}: the candidate notes are evaluated back to back, the device drifts slightly between repeats, and the repeats are noise-correlated rather than independent, so the variance reduction is weaker and each added repeat costs another calibration shift, in direct tension with the sample-efficiency constraint above.
The transferable variant of the gate is therefore the held-out chip slice: reflecting on one set of qubits/edges and judging acceptance on a disjoint set tests \emph{spatial} generalization of the note and needs no temporal freeze-and-replay, so it is realizable on hardware unchanged.
Apart from this gate, the loop, reflector, and note format are backend-agnostic: deploying on real hardware swaps the calibration backend and leaves them unchanged, with the accept gate moving from same-snapshot pairing to the held-out-slice or repeat-and-average form.

\section{Experiments}
\label{sec:experiments}

We report two results. First, the agent calibrates whole transmon chips end to end (Section~\ref{sec:exp-chip}), the prerequisite for adaptation. Second, and this is the paper's headline, an online loop specializes the agent to one persistent, drifting chip, and we validate that the learned device note is \emph{causal} by planting concrete hardware faults behind an ablation flag (Section~\ref{sec:exp-adapt}).
We then report ablation studies that isolate each agent component and the environment's physics fidelity (Section~\ref{sec:exp-ablation}).

\paragraph{Setup.}
Both the agent and reflector are \texttt{gemini-3.1-pro-preview}~\cite{geminiteam2023gemini}; the agent runs at temperature~0.4 and the reflector at temperature~0.7.
Agent runs use no API seed and are therefore non-deterministic even though the physics simulator is seeded.
Each per-qubit calibration sub-task uses a 30--40 call budget and each per-edge sub-task 40--80; difficulty tiers and noise scale are reported per experiment.

\paragraph{Experimental protocol: a four-qubit existence proof, two-qubit controlled studies.}
We demonstrate whole-chip capability once, on a four-qubit chip (Section~\ref{sec:exp-chip}); every \emph{controlled} study, namely online adaptation, planted-defect causal validation, and the agent-component and environment-fidelity ablations, is run on a two-qubit chip.
This is deliberate. Every realism effect the paper measures (CZ and Bell fidelity, gate leakage, flux distortion and its FIR predistortion, the six wall-time drift channels, and the device note itself) lives on a single edge, so a two-qubit chip isolates each effect cleanly while keeping multi-seed runs and ablation sweeps affordable.
The environment supports arbitrary qubit count $N$; the two-qubit choice is methodological isolation, not a scaling ceiling, and the four-qubit run, with wall-time-scaled drift that ages bystander qubits while their neighbors are calibrated, is the scale-up evidence.

\subsection{Whole-Chip Calibration}
\label{sec:exp-chip}

The agent calibrates a full chip by running an isolated, context-clean sub-task per qubit and per edge, accumulating results into a calibration database (Section~\ref{sec:orchestration}).
A hard-tier four-qubit grid chip (\code{noiseScale=1}), calibrated under the same configuration as the controlled studies below (per-sub-task budget 30/40, paired-three-seed gate, diagnosis reflector), ran six end-to-end calibration passes with wall-time-scaled persistent drift that ages bystander qubits while their neighbors are calibrated. Across those passes it reached a median single-qubit fidelity of $0.998$ and a median measured edge CZ fidelity of $0.887$ (range $0.81$--$0.96$), with every sub-task validly submitted; at this comfortable budget the device note was a no-op (Section~\ref{sec:exp-adapt}), as expected when the base agent is not under pressure.
A hard-tier, \code{noiseScale=2} two-qubit chip reached coverage-within-tolerance $0.833$ with measured CZ fidelity $0.986$.
These runs establish that the agent produces usable whole-chip calibrations under realistic drift and noise, which is the prerequisite for the adaptation experiment below.

Calibration quality also propagates to algorithm-level benchmarks: a calibrate-then-benchmark pipeline runs the algorithm-benchmark suite twice, once on the agent-derived noise model and once on the hidden ground truth, and the calibrated chip tracks truth with small gaps (e.g.\ XEB per-layer fidelity $0.9965$ vs.\ $0.9972$; two-qubit RB error-per-Clifford $9.9\times10^{-3}$ vs.\ $8.5\times10^{-3}$) on the runs measured so far.
These figures are single-run and illustrative; we treat the algorithm-benchmark thread as corroboration rather than a headline, and leave the discriminating version (multiple seeds and the calibration-lifetime decay curve enabled by the wall-time clock) to future work.

\subsection{Online Device Adaptation}
\label{sec:exp-adapt}

The online loop specializes the agent to \emph{one} persistent, drifting chip. Each iteration ages the device by modeled wall time, runs the incumbent and challenger notes on the \emph{same} frozen snapshot, and admits the challenger only if it clears the accept threshold (Section~\ref{sec:gate}).

\begin{figure}[h]
\centering
\includegraphics[width=\budgetFigWidth]{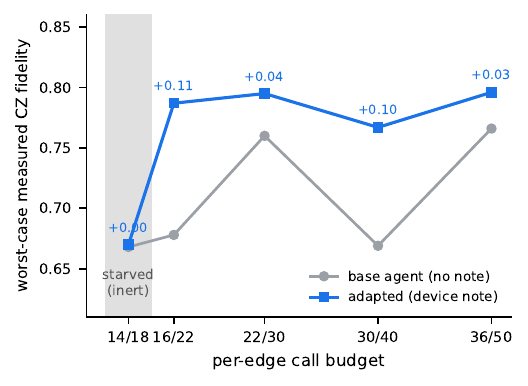}
\caption{The device note's floor-lift versus call budget (hard tier, \code{noiseScale=1}, two-qubit line, three paired seeds, six iterations per point; worst-case measured CZ fidelity over iterations, annotated with the lift). Below a minimum budget (shaded) the note is inert, because the agent has no runway to act on its advice; above it the worst-case lift is persistently positive but, at this seed count, not cleanly ordered by budget. Preliminary ($n=6$ per point).}
\label{fig:budget}
\end{figure}

\paragraph{An interpretable accepted note.}
On one hard-tier, \code{noiseScale=1} two-qubit chip run under a tight call budget (16/22 one- and two-qubit calls), which ages by modeled wall time each iteration, the loop accepts its first note at iteration~2.
On that iteration, evaluated on the same frozen snapshot, the base agent (no note) achieved measured CZ fidelity $0.678$ while the adapted agent achieved $0.913$, a $+0.235$ gain attributable to the note alone, since both ran on the identical aged device (Table~\ref{tab:adapt}).
The accepted note is interpretable device lore: it records that this chip's edge-(0,1) \code{cz_duration_us} is chronically left stale at submit and must be re-verified before submission, that its Bell fidelity is capped near $0.90$ and its iSWAP chevron is severely degraded, and that on qubit~1 the coherence times $T_2$ and $T_2^*$ are frequently left unmeasured, which correlates with reduced RB fidelity.
The first iteration proposed no accepted edit, confirming the gate rejects changes that do not clear the threshold.
This single paired accept illustrates the mechanism on one frozen snapshot; the controlled, budget-dependent version of the effect is the floor-lift reported below (Fig.~\ref{fig:budget}).

\begin{table}[h]
\centering
\small
\caption{Online adaptation on one drifting hard-tier chip (\code{noiseScale=1}, tight 16/22 budget), all six iterations. Base agent (no note) vs.\ adapted agent (device note) measured CZ fidelity on the \emph{same} frozen snapshot each iteration; the note is admitted only when it clears the accept threshold. The iteration-2 accept lifts CZ from $0.678$ to $0.913$; across all six iterations the worst-case CZ fidelity rises from $0.678$ (base) to $0.787$ (adapted).}
\label{tab:adapt}
\begin{tabular}{lccc}
\toprule
Iter & \shortstack{Base CZ\\(no note)} & \shortstack{Adapted CZ\\(note)} & Accepted \\
\midrule
1 & 0.906 & 0.906 & no (no edit) \\
2 & 0.678 & \textbf{0.913} & \textbf{yes} \\
3 & 0.795 & 0.794 & no \\
4 & 0.788 & \textbf{0.881} & \textbf{yes} \\
5 & 0.764 & \textbf{0.787} & \textbf{yes} \\
6 & 0.852 & \textbf{0.875} & \textbf{yes} \\
\bottomrule
\end{tabular}
\end{table}

\paragraph{Is the note causal? Planted-defect validation.}
A single accepted note invites a confound: does the note encode a genuine, device-specific repair, or does reflection manufacture plausible-sounding text that happens to coincide with favorable variance? We isolate causality by planting a concrete, real-hardware fault behind an ablation flag. The defect-\textsc{on} run injects a physical fault (e.g.\ a stuck tunable coupler, drive-line attenuation, or readout degradation); the matched defect-\textsc{off} control samples the device bit-for-bit identically with nothing planted. A defect-\textsc{on}/\textsc{off} pair at the same seed therefore differs \emph{only} by the planted mutation, and we ask two questions: does the note \emph{discover} the fault truth-free, and does it \emph{recover} fidelity?

The discovery claim is real, but it must be read against a baseline. On this hard-tier edge the reflector emits a generic ``iSWAP chevron fails ($r^2\approx0.2$)'' clause as boilerplate, and it appears in 9 of 10 \emph{healthy} control challengers, so the bare mention of a weak iSWAP is not by itself evidence of fault discovery. The defect-distinctive signal is the \emph{escalation} beyond that boilerplate. With a dead coupler planted on edge-(0,1) (effective coupling $g\!\times\!0.15$), the agent never sees ground truth, yet the reflector escalates to naming a \emph{missing} interaction, writing that \emph{``the interaction is heavily distorted or missing on this flux line\ldots\ prioritize CZ over iSWAP,''} in 2 of 7 challengers, language that is entirely absent (0 of 10) from the matched healthy control. The discovery claim is therefore narrow but causal: the planted fault induces a distinctive diagnosis the healthy device never produces. It is \emph{not} that the loop names every fault, since we ran the full zoo and it does not (see below), but that, for a fault it can recognize, the diagnosis is auditable device lore rather than confabulation (Fig.~\ref{fig:devicenote}).

\begin{figure}[h]
\centering
\fbox{\begin{minipage}{0.92\columnwidth}
\footnotesize\itshape
``\textbf{edge (0,1):} Chronic iSWAP failure: iSWAP fidelity consistently below $0.1$ with universally failing chevron fits ($r^2\sim0.04$--$0.12$). Do not trust iSWAP parameters; \textbf{the interaction is heavily distorted or missing on this flux line.} Bell-state fidelity also consistently poor ($0.39$--$0.53$). Coherence parameters ($T_1$, $T_2$, $T_2^*$, \upshape\texttt{drag\_beta}\itshape) are frequently left stale before submission: force re-measurement before the two-qubit tune-up, and prioritize CZ over iSWAP.''
\end{minipage}}
\caption{The device note the reflector grew on the dead-coupler-\textsc{on} chip, quoted verbatim (edge section). The agent never sees ground truth, yet the note names the planted \emph{missing} interaction (bold) from measured anomaly signatures alone; the matched healthy control never produces this missing-interaction language. This is the auditable, truth-free diagnosis that the planted-defect contrast validates.}
\label{fig:devicenote}
\end{figure}

\begin{table}[h]
\centering
\small
\caption{Planted-defect validation (preliminary; hard tier, \code{noiseScale=1}, two-qubit line, paired-three-seed gate). Measured CZ fidelity, base vs.\ adapted, on the same frozen snapshot. The dead-coupler fault is \emph{diagnosed correctly} but is physically irreparable, so the rescue magnitude is capped by the coherence ceiling the broken coupler imposes; the matched healthy control still shows the note's failure-floor lift.}
\label{tab:planted}
\setlength{\tabcolsep}{4pt}
\begin{tabular}{@{}llccc@{}}
\toprule
Arm & Iter & Base & Adapted & Names fault? \\
\midrule
Dead coupler \textsc{on} & 1 & 0.417 & 0.471 & \textbf{yes} \\
Healthy \textsc{off} & 1 & 0.766 & 0.909 & no \\
\bottomrule
\end{tabular}
\end{table}

\paragraph{What the magnitude does and does not show.}
The recovery magnitude must be read carefully (Table~\ref{tab:planted}). A dead coupler is \emph{diagnosed} perfectly but is physically irreparable, since no in-vocabulary workaround can restore a coupling that is not there, so its fidelity rescue is small ($+0.05$) and bounded by physics, not by the note. Conversely the matched healthy control is not silent on \emph{fidelity}: from a low-base iteration the note still lifts CZ fidelity ($0.766\!\to\!0.909$), consistent with our broader finding that \textbf{the note raises the failure floor and cuts variance rather than the mean}. It is failure-insurance that pays off precisely when the base agent is failing. We therefore report floor, worst-case, and variance reduction rather than mean fidelity, and we separate the two claims cleanly: planted defects establish \emph{causal diagnosis}, and the floor-lift establishes \emph{reliability value}. The headline is that the loop \emph{diagnoses} a specific planted fault truth-free; the recovery magnitude is a separate, harder bar, which we address next.

\paragraph{No clean defect-specific recovery case exists.}
We ran the full twelve-fault zoo (hard tier, \code{noiseScale=1}, two-qubit line, paired-three-seed gate, six iterations per fault) specifically to find a defect whose workaround is in the agent's vocabulary and would therefore show a real fidelity \emph{rescue}. None did, for a structural reason. The base agent runs a full fresh calibration on every attempt, so any fault whose remedy is simply to re-measure a parameter is already corrected by the baseline before the note can contribute. Drive-line attenuation ($\times0.55$ drive, corrected by re-deriving \code{pi_amp} from Rabi), LO instability, charge jumps, and off-sweet-spot drift (all corrected by re-measuring the frequency) came back \emph{null}: the defect-\textsc{on} base fidelity matched the healthy control (for instance, \code{drive_attenuation} \textsc{on} reaches Bell mean $0.831$ against control $0.808$), and the note added no defect-specific rescue. The faults that \emph{survive} a fresh calibration are physical ones a re-measurement confirms but cannot fix, such as a dead coupler or a TLS collision that collapses $T_1$, and these are diagnosed but irreparable. The tension is intrinsic: where the note can diagnose, the baseline can usually already fix, and where the baseline fails, the fault is irreparable. We therefore do not claim a defect-specific recovery. The more important claim, which the planted defects and the floor-lift jointly support, is that the device note is a truth-free \emph{diagnosis} and \emph{failure-insurance} mechanism, naming device-specific faults and raising the worst-case fidelity floor, not a defect-specific repair tool.

\paragraph{When does the note pay off?}
Because the note's value is failure-insurance, it should matter most when the base agent is under pressure. We swept the per-sub-task call budget on the same quiet device (hard tier, \code{noiseScale=1}, no planted defect, three paired seeds, six iterations per point) from \emph{starved} ($14/18$ one-/two-qubit calls) to \emph{ample} ($36/50$). The clearest finding is a \emph{threshold} (Fig.~\ref{fig:budget}): at the starved budget the floor-lift vanishes ($+0.002$ worst-case CZ), because the agent has no runway to act on the note's advice; above the threshold the worst-case lift is persistently positive ($+0.03$ to $+0.11$) but without clean budget-dependence at this seed count, and, against a naive coherence-ceiling argument, the ample budget did \emph{not} null it. What we can claim is therefore the threshold effect: the note needs a minimum operating budget to pay off, below which it is inert; a sharper budget-response curve needs more seeds.
This floor-lift also scales to the whole four-qubit chip: under the same tight budget the worst-case measured CZ fidelity rises from $0.700$ without the note to $0.810$ with it over six iterations, the budget-pressure floor-lift at chip scale rather than a defect-specific repair (the drive fault planted on that chip is absorbed by the agent's fresh per-attempt calibration, as discussed above).

\subsection{Ablation Studies}
\label{sec:exp-ablation}

The environment's ablation flags (Section~\ref{sec:environment}) expose each agent component and each physics layer independently. We study them on a single hard-tier two-qubit edge, with the seed count reported per study (Section~\ref{sec:experiments}, Setup). One edge episode exercises every agent component and every edge-level realism effect, so this configuration isolates each factor in turn while keeping the sweeps tractable; chip-scale ablation sweeps are left to future work.

\paragraph{Agent-component ablations.}
Each agent component is exposed as an additive default-on flag, toggled independently while holding the device fixed. We report the effect (relative to the full agent) on coverage-within-tolerance and measured Bell fidelity when each is disabled:
(i)~\textbf{vision}, with the rendered measurement plots withheld from the function responses (text and scalars only), a first, naive probe of the ``vision-language'' framing that we sharpen below;
(ii)~\textbf{notebook injection}, with the per-turn authoritative parameter block removed;
(iii)~\textbf{history compression}, with measurement turns left uncompressed (longer context, stale plots retained);
(iv)~\textbf{fit-quality gating}, with the $r^2$ trust rule removed from the prompt (we expect coverage to fall as the agent commits noise-dominated fits, which is the point);
(v)~\textbf{runtime nudges}, with the supervisor guardrails disabled.

Toggled individually at fifteen seeds on one hard-tier two-qubit edge with no parameters pre-loaded, no single component moves measured Bell fidelity: the median sits at $0.90$--$0.91$ across all six arms and the worst-case floor stays flat (Table~\ref{tab:component}), with only within-noise coverage differences and a small coverage cost for removing the notebook or history compression. Each component is therefore robustness machinery, preventing empty episodes, budget exhaustion, and noise-dominated commits, rather than a lever on measured fidelity. This holds under budget pressure too: re-running the full toggle set at a tight $35$-call budget, the regime in which the device note's value appears (Section~\ref{sec:exp-adapt}), leaves the null intact, with \code{no_vision} Bell $0.908$ against the full agent's $0.912$ and no component ordering in the floor.
The plain vision toggle is null for a specific and informative reason: the environment mirrors every plot annotation into text (detected peaks and TLS candidates, fitted values, recommended settings, one-line evidence summaries), so withholding the rendered figure removes pixels but not information. That toggle therefore does not test whether the plot itself carries information the text lacks; the next experiment does.

\begin{table}[h]
\centering
\small
\caption{Agent-component ablations on one hard-tier two-qubit edge (fifteen seeds, no pre-loading). Median and worst-case (floor) Bell fidelity measured on the truth device. No single toggle moves measured fidelity; coverage differences are within rollout noise. The plain \code{no_vision} row is null because the tools mirror every plot into text, a redundancy the scalar-deprived test (Table~\ref{tab:vision}) removes.}
\label{tab:component}
\begin{tabular}{lcc}
\toprule
Ablation & Bell (median) & Bell floor \\
\midrule
Full agent      & $0.902$ & $0.792$ \\
no vision       & $0.909$ & $0.823$ \\
no notebook     & $0.910$ & $0.817$ \\
no compression  & $0.907$ & $0.824$ \\
no fit-gating   & $0.909$ & $0.811$ \\
no nudges       & $0.910$ & $0.785$ \\
\bottomrule
\end{tabular}
\end{table}

\paragraph{The plot carries information once the text mirror is closed.}
To test whether the rendered plot itself carries information, we close the mirrored text channels: measurement and fit responses are pruned to fit-quality and warnings, and value-carrying evidence summaries are blanked, so a parameter value appears \emph{only} in the plot. We then compare three arms over thirty seeds on full, no-preload episodes (Table~\ref{tab:vision}): \emph{full} (vision and scalars), \emph{plot-only} (scalars stripped, plot kept), and \emph{blind} (both stripped). The result is monotonic on parameter recovery: coverage $0.824$ (full) $>$ $0.562$ (plot-only) $>$ $0.402$ (blind). The plot alone is thus worth $+0.160$ coverage over the blind agent, a roughly $40\%$ relative lift, and it eliminates catastrophic failure: $4$ of $30$ blind episodes recover zero parameters, none in the plot-only or full arms. Consistent with every result here, the value lands on parameter recovery, not on coherence-limited mean Bell, which does not track coverage: the plot-only arm's mean Bell ($0.772$) is not above the blind arm's ($0.832$). The precise claim is therefore that the multimodal plot channel is load-bearing for parameter recovery and for avoiding catastrophic blind failures, and is not a lever on physics-limited gate fidelity. This also explains the earlier null (Table~\ref{tab:component}): the plain vision toggle sees nothing only because the environment redundantly mirrors the plots into text; closing that mirror reveals the plot's real contribution.

\begin{table}[h]
\centering
\small
\caption{Scalar-deprived vision test (thirty seeds, hard two-qubit edge, no pre-loading, $80$-call budget). Arms: \emph{full} (vision and scalars), \emph{plot-only} (scalars stripped, plot kept), \emph{blind} (both stripped); with the text channels closed a parameter value appears only in the plot. Coverage (mean parameter recovery) is monotonic in visual access; the plot alone is worth $+0.16$ coverage over blind. The last column counts episodes that recover zero parameters. Mean Bell is coherence-limited and does not track coverage.}
\label{tab:vision}
\begin{tabular}{lccc}
\toprule
Arm & Coverage & Bell & cov $=0$ \\
\midrule
Full        & $0.824$ & $0.882$ & $0/30$ \\
Plot only   & $0.562$ & $0.772$ & $0/30$ \\
Blind       & $0.402$ & $0.832$ & $4/30$ \\
\bottomrule
\end{tabular}
\end{table}

\textbf{Environment fidelity vs.\ a toy environment.}
The central claim of contribution~1 is that the \emph{whole} environment, not merely its noise model, is physics-grounded. We test this by running the identical agent (all components on) on a matched pair of two-qubit edges that differ \emph{only} in environment-physics fidelity: the full hard-tier environment, and a \emph{toy environment} with every physics layer stripped (flat cosine $f_{01}(\Phi)$ in place of the circuit-quantization truth, an identity flux line with no distortion, no CZ leakage, and all six drift channels off), while the scalar truth ($g$, $\zeta$, $T_1$, $T_2$, $f_{01}$) and the readout model are held fixed. The toy environment is the conventional $T_1/T_2$ calibration simulator one would build without these deliverables, not a strawman depolarizing channel.

The two registers (Table~\ref{tab:envfid}) tell a consistent story. First, the proxy a toy environment reports is \emph{blind} to the gap: parameter coverage within tolerance ($0.857$) and 1Q RB ($0.999$, shared coherence) are identical across the two worlds. Second, the physics-grounded environment exposes a measured-fidelity gap that is otherwise invisible: Bell fidelity $0.911$ vs.\ $0.985$ and iSWAP fidelity $0.65$ vs.\ $0.90$, with a flux-distortion residual of $0.0158$ that is identically zero in the toy world. Third, the agent's advanced machinery is \emph{inert} in the toy environment: it fires FIR predistortion in both worlds ($\sim$3 calls per episode), but with nothing to correct the committed residual is exactly zero, so a toy environment would ``validate'' calibration strategies (FIR predistortion, leakage-aware CZ) that are never actually exercised, precisely the strategies that must transfer to hardware. A corroborating signature: toy-environment Bell fidelity is deterministic to five digits across all eight seeds (variance collapse), while the full environment carries real seed-to-seed spread. This is the aggregate physics gap; per-layer attribution (the residual cleanly isolates flux distortion) is left to the ablation ladder in future work.

\begin{table}[t]
\centering
\small
\caption{Environment-fidelity ablation: identical agent on one hard-tier two-qubit edge (eight seeds, medians). The toy environment strips every physics layer (circuit truth, flux distortion, CZ leakage, drift) but keeps $T_1/T_2$ and the readout model. Coverage and 1Q RB are blind to the gap; measured two-qubit fidelity and the flux-distortion residual are not.}
\label{tab:envfid}
\begin{tabular}{lrr}
\toprule
Metric & \multicolumn{1}{c}{Full env.} & \multicolumn{1}{c}{Toy env.} \\
\midrule
Coverage within tolerance   & $0.857$  & $0.857$ \\
1Q RB avg.\ fidelity        & $0.999$  & $0.999$ \\
Bell fidelity               & $0.911$  & $0.985$ \\
iSWAP fidelity              & $0.650$  & $0.899$ \\
Predistortion residual RMS  & $0.0158$ & $0.0000$ \\
CZ chevron peak             & $0.900$  & $0.901$ \\
FIR-predistortion calls     & $2.6$    & $3.2$ \\
\bottomrule
\end{tabular}
\end{table}

\textbf{FIR vs.\ single-pole predistortion.}
On 8 hard-tier device seeds using a dense cryoscope grid (100 markers over 300~ns, 12\,000 shots per marker), the 100-tap FIR filter ($\Delta t = 4$~ns, Tikhonov deconvolution) achieves a mean predistortion residual RMS of $0.0106 \pm 0.0030$ versus $0.0125 \pm 0.0028$ for the single-pole fit, a $\sim$15\% improvement, with FIR winning on all 8 seeds.
Both methods satisfy the scoring tolerance of $0.04$; the gap is most pronounced on devices whose flux kernel includes an AWG LC ring, which a single exponential cannot approximate.

\textbf{Empty-note vs.\ adapted-note baseline.}
Because every online iteration evaluates the base agent (empty note) and the adapted agent on the same frozen snapshot, the empty-vs-adapted comparison is read directly off the per-iteration logs and reported above (Tables~\ref{tab:adapt},~\ref{tab:planted}); a hand-scripted expert-recipe baseline is deferred to future work.

\section{Discussion and Limitations}
\label{sec:discussion}

\paragraph{What is realistic.}
The environment captures the end-to-end structure of a calibration shift followed by chip benchmarking: hidden chip truth, an agent-derived calibration database, a calibrated noise model and a truth noise model, and the benchmark deltas between them.
The realism features are physically grounded rather than synthetic stress tests bolted on.
Device truth comes from circuit quantization, flux distortion is applied by true LTI convolution at current-hardware magnitudes, drift is paced by a wall-clock and persists across a chip, and gate quality is \emph{measured} on the truth device with calibrated parameters.
Each feature is independently ablatable, which is what makes the environment a controlled instrument rather than a single opaque difficulty knob.

\paragraph{What is synthetic.}
The environment is not a hardware digital twin.
Flux noise, TLS spectral diffusion, leakage, and drift are physically motivated models, not calibrated telemetry from a specific refrigerator, and the flux-distortion magnitudes are taken from a published hardware regime rather than measured in-house~\cite{venkateswaran2026dpd}.
The calibrated noise model collapses the device into compact circuit-noise parameters and does not preserve full TLS-bath dynamics, correlated readout, pulse-distortion history, queue latency, or instrument failure modes.
The simulator is seeded, but agent runs are not deterministic, because the model calls use nonzero temperature and no API seed; the results in Section~\ref{sec:experiments} are therefore single- or few-seed and must be repeated with error bars, which we flag throughout.

\paragraph{On the adaptation claims.}
The online results are promising but preliminary, and we are deliberate about what they do and do not show.
The strongest single-run evidence is the paired-snapshot result, a $+0.235$ measured CZ-fidelity gain (from $0.678$ to $0.913$) attributable to a device-note edit alone, because the paired evaluation structurally isolates strategy from drift.
The strongest \emph{causal} evidence is the planted-defect contrast: with a fault injected behind an ablation flag, the reflector's note names the fault truth-free, which the matched no-defect control does not, separating genuine device diagnosis from confabulation.
Two caveats follow from those runs.
First, a perfectly diagnosed but physically irreparable fault such as a dead coupler yields only a small fidelity rescue, because no in-vocabulary workaround can restore missing coupling.
Second, having run the full fault zoo, we found that a clean defect-specific \emph{recovery} case does not exist: faults whose workaround is to re-measure a parameter are already corrected by the agent's fresh per-attempt calibration, while the faults that survive a fresh calibration are precisely the irreparable ones (Section~\ref{sec:exp-adapt}).
We therefore characterize the note as diagnosis plus failure-insurance, not defect-specific repair.
Its effect is best described as raising the failure \emph{floor} and cutting variance rather than the mean, so we report worst-case and variance rather than mean fidelity to avoid overclaiming, and we find that this floor-lift requires a minimum call budget to materialize at all.
By design the loop \emph{overfits} to one chip; the open question we want to quantify is how quickly the accept gate converts that overfitting into a measurable, drift-robust gain.

\paragraph{Truth leakage and isolation.}
The agent isolates calibration knowledge in the system prompt and never sees ground truth, and the online reward is structurally truth-free.
The remaining leakage is in chip-level task strings, which still mention measurement strategy in places; these should be reduced to required outputs, leaving measurement choices to the agent interface.

\paragraph{Toward deployment.}
The artifacts are co-designed for one endpoint: running this agent on a real superconducting transmon chip.
The tool interfaces mirror real control-electronics primitives, the truth-free online reward is computable on a real device, the device note is auditable and rollback-safe, and a backend shim that swaps the simulator for hardware measurement leaves the loop, reflector, and device note unchanged.
The one component that does not transfer verbatim is the same-snapshot accept gate.
Freezing and replaying an identical drifted snapshot is a simulation-only affordance (Section~\ref{sec:gate}), so on hardware the gate moves to its held-out-chip-slice form (spatial generalization, no temporal replay) or to repeat-and-average, the latter trading sample efficiency for variance reduction under residual drift.
The environment is best read not as a claim that a model should run hardware unsupervised, but as a controlled laboratory for the experimental policy and the gradient-free, interpretable specialization that such a deployment would require.

\bibliographystyle{unsrtnat}
\bibliography{references}

\appendix
\section{Appendix}
\label{sec:appendix-plots}

\subsection{Representative Tool-Call Plots}

Section~\ref{sec:agent} states that the agent's primary evidence is a rendered plot, not a scalar summary; this appendix shows what that evidence actually looks like.
All panels are drawn from hard-tier two-qubit chip-pipeline runs, none from an idealized or noise-free configuration.
The intent throughout is to make the noise floor, drift artifacts, and fit-quality judgments in Section~\ref{sec:environment} concrete rather than descriptive.

\begin{figure}[htbp]
\centering
\includegraphics[width=0.92\linewidth]{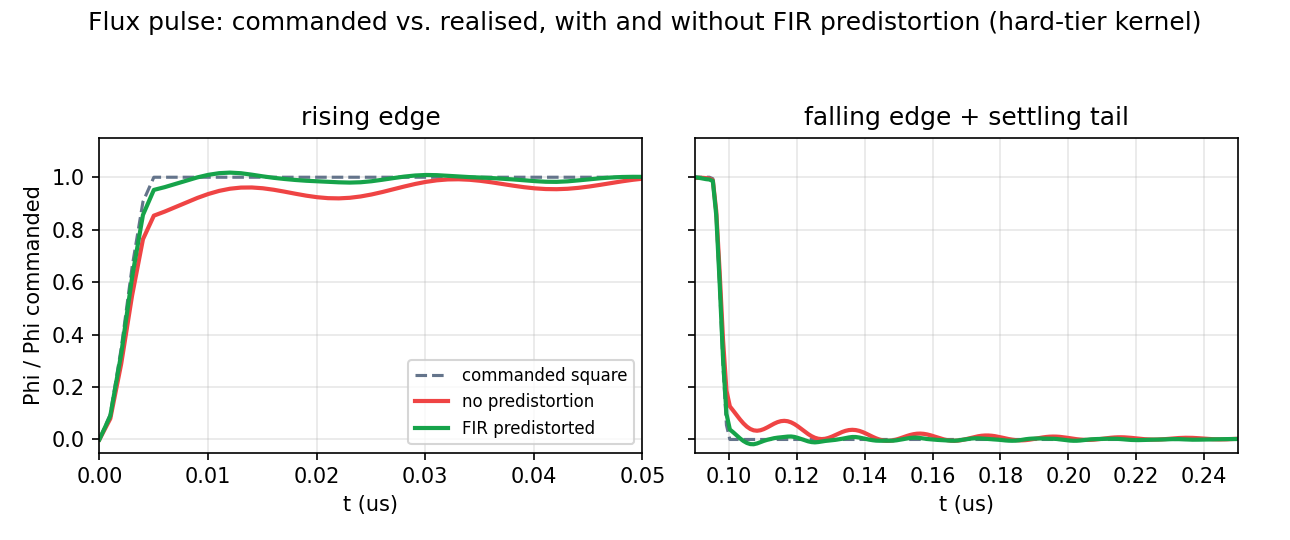}
\caption{Pulse-level view of a hard-tier coupler flux-line kernel of the same form as the one behind the Fig.~\ref{fig:app-1q}(d) cryoscope (a two-pole $+$ bias-tee $+$ AWG-ring transfer function), rebuilt from the raw commanded/realized traces the dispatcher saves alongside every cryoscope call, at the point in a run where the agent's committed FIR predistortion achieved its best fit (RMS residual $0.029\rightarrow 0.008$). The commanded pulse (dashed) is a $0.1$~$\mu$s square; without predistortion (red) the realized pulse overshoots on the rising edge and rings for tens of nanoseconds after the falling edge; with the FIR correction (green) both edges track the command more closely. This is the pulse-level counterpart of the step-response ratio in the cryoscope panel, and the effect predistortion is scored against (Section~\ref{sec:flux}).}
\label{fig:app-flux}
\end{figure}

\begin{figure}[htbp]
\centering
\begin{subfigure}[b]{0.48\linewidth}
  \includegraphics[width=\linewidth]{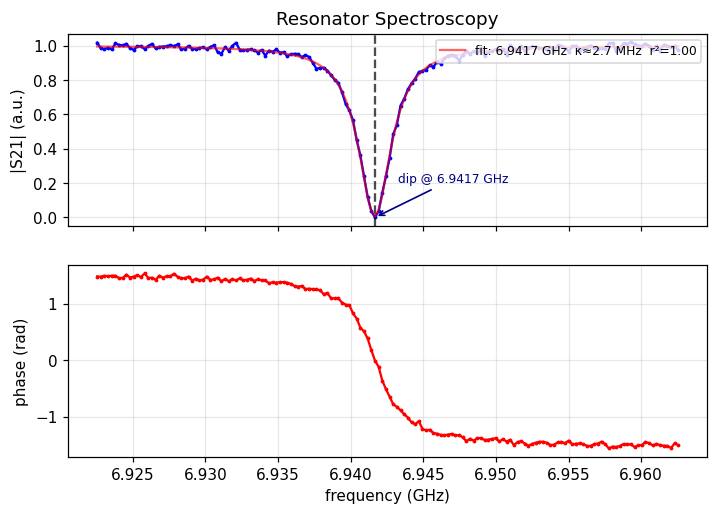}
  \caption{Resonator spectroscopy}
  \label{fig:app-resonator}
\end{subfigure}
\hfill
\begin{subfigure}[b]{0.48\linewidth}
  \includegraphics[width=\linewidth]{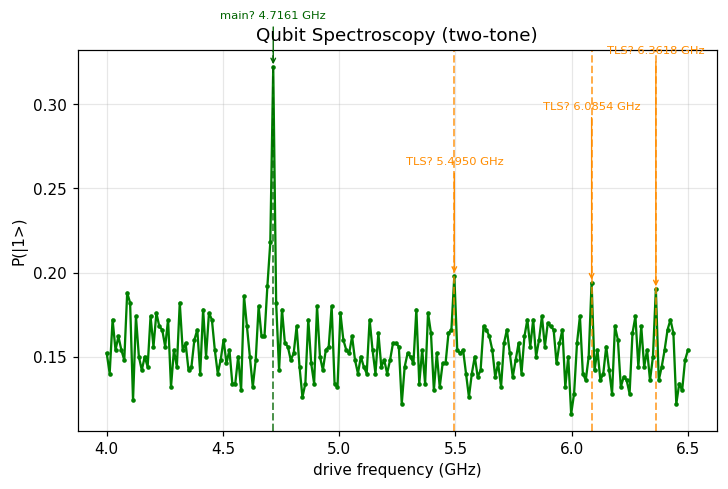}
  \caption{Qubit spectroscopy}
  \label{fig:app-qubitspec}
\end{subfigure}
\vspace{0.4em}

\begin{subfigure}[b]{0.48\linewidth}
  \includegraphics[width=\linewidth]{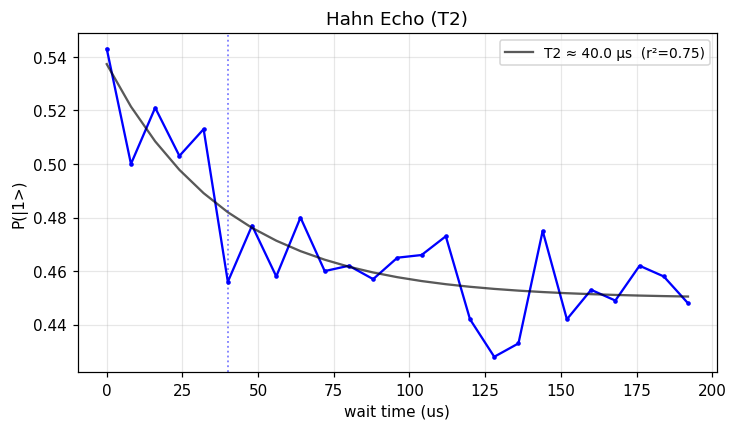}
  \caption{Hahn echo}
  \label{fig:app-echo}
\end{subfigure}
\hfill
\begin{subfigure}[b]{0.48\linewidth}
  \includegraphics[width=\linewidth]{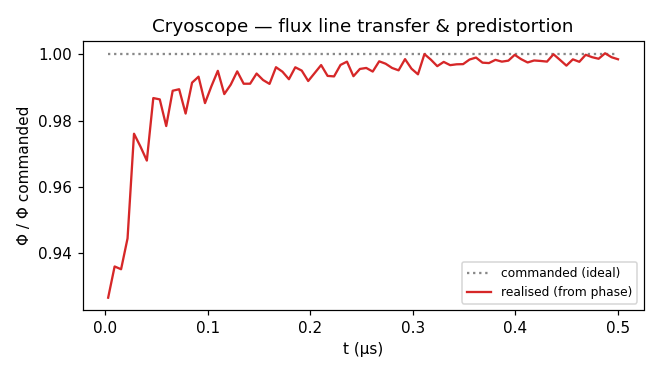}
  \caption{Cryoscope}
  \label{fig:app-cryoscope}
\end{subfigure}
\caption{Single-qubit tool-call plots, hard tier. (a) Resonator spectroscopy, narrow zoom: once the coarse dip location is known, a $40$~MHz local rescan resolves it cleanly ($\kappa\approx 2.7$~MHz, $r^2{=}1.00$), the kind of confirmatory high-resolution scan a coarse-then-fine strategy relies on. (b) Qubit spectroscopy on q1 (drive amp $0.1$, $500$ shots): a coarse $2.5$~GHz locate scan finds one narrow, sharp main peak at $4.716$~GHz, later confirmed at $4.712$~GHz, alongside three algorithmically-flagged ``TLS?'' candidates ($5.495$, $6.085$, $6.362$~GHz) that are barely above the noise floor and far detuned; the agent locked onto the correct peak and did not chase the other three. (c) Hahn echo: the $T_2$ fit ($r^2 = 0.75$) is visibly noisy at long wait times, the kind of borderline fit the FIT QUALITY protocol (Section~\ref{sec:agent}) asks the agent to flag rather than commit outright. (d) Cryoscope: the realized-to-commanded flux ratio recovers the bulk of its deviation within the first tens of nanoseconds (the fast dominant pole of Section~\ref{sec:flux}) and then settles the residual toward $1.0$ over hundreds of nanoseconds with visible ringing, the slow tail and ring of the flux-line transfer function that predistortion must invert.}
\label{fig:app-1q}
\end{figure}

\begin{figure}[htbp]
\centering
\begin{subfigure}[b]{0.48\linewidth}
  \includegraphics[width=\linewidth]{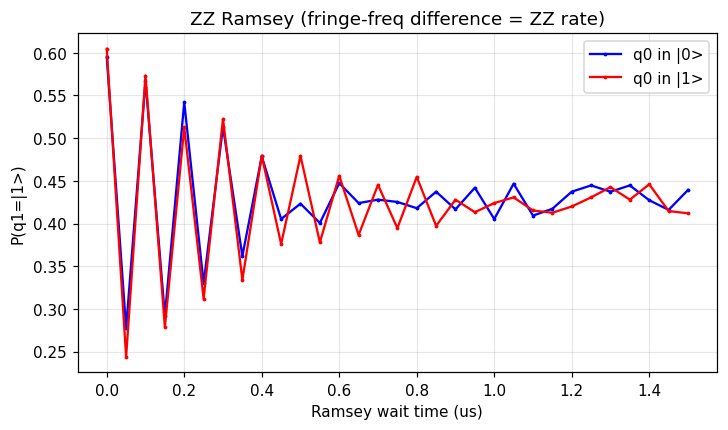}
  \caption{$ZZ$ Ramsey}
  \label{fig:app-zzramsey}
\end{subfigure}
\hfill
\begin{subfigure}[b]{0.48\linewidth}
  \includegraphics[width=\linewidth]{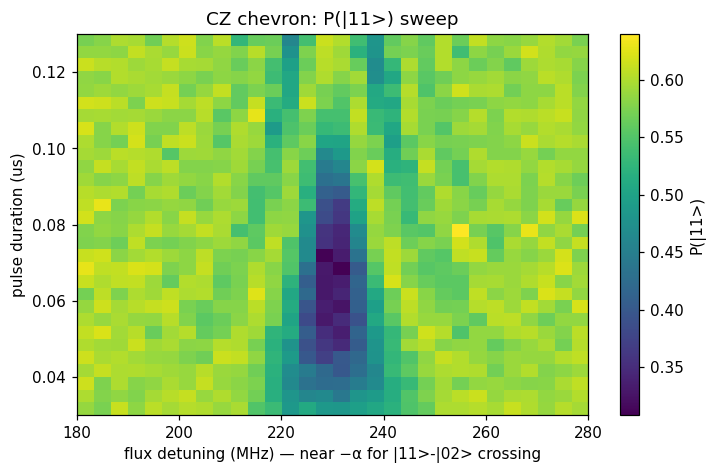}
  \caption{CZ chevron}
  \label{fig:app-czchevron}
\end{subfigure}
\vspace{0.4em}

\begin{subfigure}[b]{0.48\linewidth}
  \includegraphics[width=\linewidth]{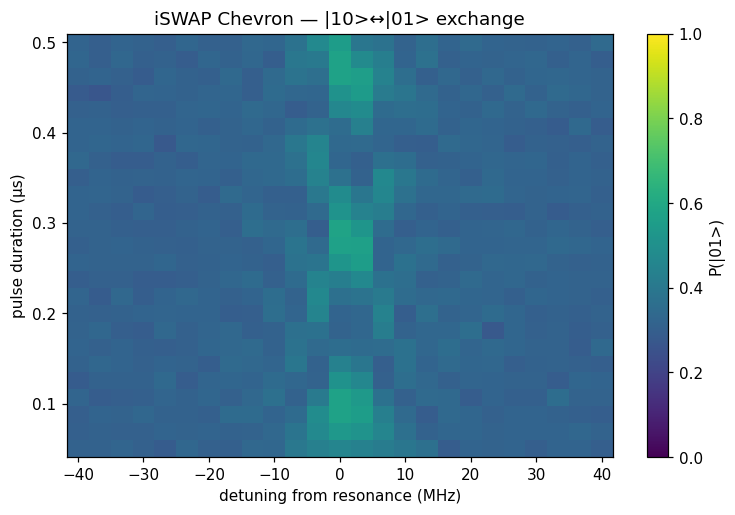}
  \caption{iSWAP chevron}
  \label{fig:app-iswapchevron}
\end{subfigure}
\hfill
\begin{subfigure}[b]{0.48\linewidth}
  \includegraphics[width=\linewidth]{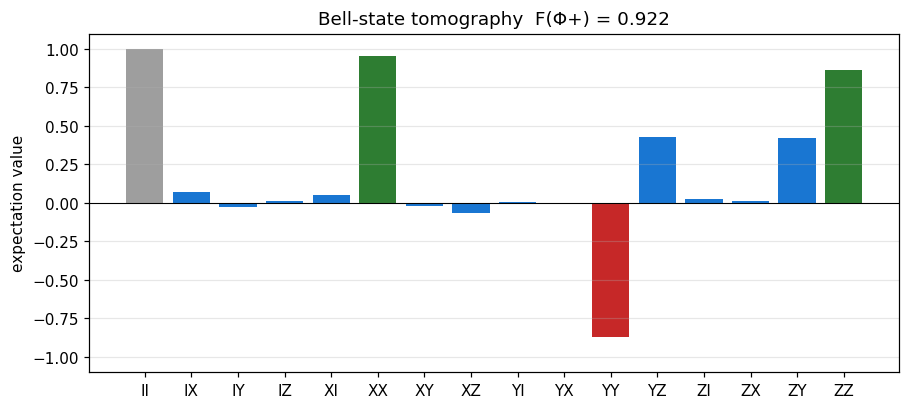}
  \caption{Bell-state tomography}
  \label{fig:app-bell}
\end{subfigure}
\caption{Two-qubit tool-call plots, hard tier. (a) $ZZ$ Ramsey: the two spectator traces (q0 in $|0\rangle$ vs.\ $|1\rangle$) share a decay envelope but differ in fringe frequency; the shared-phase joint fit (Section~\ref{sec:environment}) extracts $\zeta$ from that frequency difference rather than either trace alone. (b) CZ chevron: the avoided-crossing structure near the $|11\rangle\!\leftrightarrow\!|02\rangle$ resonance is visible but pixel-noisy, which is why the 2D joint fit replaced the earlier FFT-of-1D-cut approach. (c) iSWAP chevron: the exchange fringe is present but low-contrast under drift and leakage, closer to a real chevron scan than a clean textbook figure. (d) Bell-state tomography: most Pauli expectation values are near zero as expected, but off-diagonal terms ($XI$, $XZ$, $YZ$, $ZX$) are visibly nonzero, the residual coherent error the fidelity number $F=0.922$ alone does not show.}
\label{fig:app-2q}
\end{figure}

\subsection{Representative Decision Traces}
\label{sec:app-traces}

The plots above are the kind of evidence the agent acts on; this section shows how it reasons over such evidence.
Before each tool call the agent emits a short reasoning log (``I know / I need / Next''), described in Section~\ref{sec:orchestration}, that records the belief driving the call.
Tables~\ref{tab:trace-1q} and \ref{tab:trace-2q} reproduce two abridged traces from hard-tier chip-pipeline runs, quoting that log verbatim next to the tool called and the commit or rejection that followed.
The plots in Figures~\ref{fig:app-1q} and \ref{fig:app-2q} and these traces are separate representative examples of the same hard-tier pipeline, not two views of one episode.
The logs are the agent's self-reported justifications rather than a readout of its internal computation, but they are the same records used for the notebook and for after-the-fact auditing, and they make the coarse-then-fine, fit-gating, and guard-fallback behaviors of Section~\ref{sec:agent} concrete.

\begin{table}[htbp]
\centering
\small
\caption{Single-qubit decision trace (q0), abridged. Two behaviors are visible: the agent treats a coarse spectroscopy peak as a candidate rather than a commit and confirms it with a narrow power-broadened rescan, the same TLS disambiguation illustrated in Fig.~\ref{fig:app-qubitspec}; and it declines a borderline echo fit and re-measures on a tighter window before committing $T_2$, the fit-quality gate of Section~\ref{sec:agent} acting on a borderline fit like that in Fig.~\ref{fig:app-echo}. Intermediate steps are elided ($\cdots$).}
\label{tab:trace-1q}
\begin{tabular}{@{}p{0.24\linewidth}p{0.70\linewidth}@{}}
\toprule
Tool call & Agent reasoning log (verbatim) and outcome \\
\midrule
\code{run_resonator_spectroscopy} & ``The device is uncalibrated. I need the readout resonator frequency to anchor all subsequent measurements. Next: a wide sweep from 6.5 to 7.5 GHz.'' $\Rightarrow$ dip at $6.942$~GHz, commit \code{resonator_freq_GHz}. \\
\addlinespace
\code{run_qubit_spectroscopy} (coarse) & ``The coarse qubit spectroscopy shows a peak at $5.163$~GHz, but it could be a TLS. I need to verify with a narrow sweep and higher drive amplitude to power-broaden the real peak.'' $\Rightarrow$ \emph{no commit}. \\
\addlinespace
\code{run_qubit_spectroscopy} (narrow, drive$=0.3$) & ``A clear peak at $5.163838$~GHz with $r^2{=}0.975$.'' $\Rightarrow$ commit \code{qubit_freq_GHz}. \\
\midrule
\multicolumn{2}{@{}l}{$\cdots$ Rabi, Ramsey (with one \code{revise_param} on \code{qubit_freq_GHz}), $T_1$ $\cdots$} \\
\midrule
\code{run_echo} (window $400~\mu$s) & ``$T_2$ is approximately $22~\mu$s, but the fit was poor ($r^2{=}0.618$) because the window was too wide. I need a better $T_2$ fit.'' $\Rightarrow$ \emph{no commit}, re-run on a tighter window. \\
\addlinespace
\code{run_echo} (window $200~\mu$s) & ``$T_2$ is $28.074~\mu$s with a good fit.'' $\Rightarrow$ commit \code{T2_us}. \\
\bottomrule
\end{tabular}
\end{table}

\begin{table}[htbp]
\centering
\small
\caption{Two-qubit-edge decision trace, abridged. The last two rows show the guarded-fallback behavior of Section~\ref{sec:agent}: the iSWAP 2D chevron fit fails the quality gate on a low-contrast scan of the kind in Fig.~\ref{fig:app-iswapchevron}, the tool nulls its recommendation and returns a warning, and rather than commit a garbage value the agent falls back to the documented physical prior $1/(2g)$ computed from the already-committed coupling.}
\label{tab:trace-2q}
\begin{tabular}{@{}p{0.24\linewidth}p{0.70\linewidth}@{}}
\toprule
Tool call & Agent reasoning log (verbatim) and outcome \\
\midrule
\code{run_zz_ramsey} & ``ZZ Ramsey gave \code{zz_rate_MHz}${=}0.1994$ and \code{coupling_g_MHz}${=}6.5847$ with $r^2{=}0.97$.'' $\Rightarrow$ commit both. \\
\addlinespace
\code{run_cryoscope} $+$ \code{fit_step_response} & ``The step response fit gave $\tau{=}0.1271~\mu$s and distortion amp $-0.1292$.'' $\Rightarrow$ commit predistortion before the CZ chevron. \\
\addlinespace
\code{run_cz_chevron} & ``The CZ chevron gave \code{recommended_cz_duration_us}${=}0.1270$ and detuning $230.1$~MHz.'' $\Rightarrow$ commit \code{cz_duration_us}. \\
\addlinespace
\code{run_bell_tomography} & ``Bell fidelity is $0.8929$.'' $\Rightarrow$ commit \code{bell_state_fidelity}. \\
\addlinespace
\code{run_iswap_chevron} & 2D fit returns $r^2{=}-1.93 < 0.4$; the tool rejects its recommendation and warns. \\
\addlinespace
(fallback) & ``The iSWAP chevron fit failed, so I will compute \code{iswap_duration_us}${=}1/(2g){=}0.0759~\mu$s.'' $\Rightarrow$ commit from the physical prior. \\
\bottomrule
\end{tabular}
\end{table}

\subsection{The Compressed Calibration Notebook}
\label{sec:app-notebook}

Section~\ref{sec:agent} describes the calibration notebook as the compact, authoritative state the agent reasons over, and the history compression that keeps plot-heavy transcripts inside the context window.
Figure~\ref{fig:app-notebook} shows a real instance: the \code{=== CALIBRATION NOTEBOOK ===} block that is re-rendered and re-injected on every turn, captured verbatim from a hard-tier two-qubit edge sub-episode of a chip-pipeline run.
It is a separate representative example, not the same episode as the plots or traces above.
Each set parameter carries its committed value and a one-line evidence record (the tool that set it and its fit quality $r^2$); the header reports the elapsed modeled wall time, and the wall-time drift-age tags (\code{[age:N min]}, escalating to \code{drift?}) flag values old enough to warrant a re-check.
Two behaviors from Section~\ref{sec:agent} are visible in the state itself.
First, the coherence and DRAG values carried over from the single-qubit stage (\code{prior_calibration}) are flagged \code{[STALE]} once an upstream re-measurement invalidated them, so the notebook is telling the agent to re-measure $T_1$, $T_2$, $T_2^*$, and \code{drag_beta} before trusting the two-qubit tune-up; this is the ``coherence times frequently left unmeasured'' pattern the accepted device note learns to guard against (Section~\ref{sec:exp-adapt}).
Second, the iSWAP chevron fit failed its quality gate ($r^2{=}-0.07$, \code{recommended=null}), so rather than commit a garbage value the agent fell back to the physical prior $1/(2g)$, which for the committed $g{=}13.25$~MHz gives the $0.0377~\mu$s recorded for \code{iswap_duration_us}, the guarded-fallback behavior of Table~\ref{tab:trace-2q}.
Below the block, history compression rewrites a committed measurement turn, dropping its JSON payload and rendered PNG, to a single marker of the form \code{[compressed:run_rabi step=8 -> Rabi first-max amp 0.700]} that preserves the one fact the notebook already records.

\begin{figure}[htbp]
\centering
\begin{Verbatim}[frame=single,framesep=5pt,fontsize=\scriptsize,baselinestretch=0.95]
=== CALIBRATION NOTEBOOK (elapsed: 12.2 min) ===
[stage 0] resonator_freq_GHz  = 7.3481   [step 0, prior_calibration]
[stage 0] qubit_freq_GHz      = 4.695    [step 6, run_qubit_spectroscopy, r^2=0.98]
[stage 2] pi_amp              = 0.2827   [age:7min] [step 8, run_rabi, r^2=1.00]
[stage 2] pi_duration_us      = 0.04     [step 9, run_rabi, r^2=1.00]
[stage 2] pi2_amp             = 0.1414   [step 29, run_rabi, r^2=1.00]
[stage 2] pi2_duration_us     = 0.04     [step 30, run_rabi, r^2=1.00]
[stage 2] drag_beta           = 0.397    [STALE] [age:12min] [step 0, prior_calibration]
[stage 3] T1_us               = 23.676   [STALE] [step 0, prior_calibration]
[stage 3] T2_us               = 21.141   [STALE] [step 0, prior_calibration]
[stage 3] T2star_us           = 1.384    [STALE] [step 0, prior_calibration]
[stage C] zz_rate_MHz         = 1.1778   [step 12, run_zz_ramsey, r^2=0.99]
[stage C] coupling_g_MHz      = 13.2547  [step 13, run_cz_chevron, r^2=0.92]
[stage D] cz_duration_us      = 0.05335  [step 14, run_cz_chevron: rec_cz=0.0533us, r^2=0.92]
[stage E] bell_state_fidelity = 0.8597   [step 21, run_bell_tomography: F=0.8597]
[stage E] cz_gate_fidelity    = 0.9064   [step 22, run_bell_tomography: F=0.8597]
[stage F] crosstalk_amp       = 0.0951   [step 23, run_crosstalk: ratio=0.0951]
[stage G-iswap]      iswap_duration_us   = 0.03772  [step 24, run_iswap_chevron: rec=null]
[stage G-iswap]      iswap_gate_fidelity = 0.9      [step 25, run_iswap_chevron: rec=null]
[stage G-predist]    predistortion_tau_us = 0.11578 [step 27, fit_step_response, r^2=1.00]
[stage G-predist]    predistortion_amp    = 0.062   [step 28, fit_step_response, r^2=1.00]
[stage G-flux-xtalk] flux_crosstalk_amp   = 0.0382  [step 26, run_joint_flux_sweetspot]
unset: allxy_pass, avg_gate_fidelity, readout_fidelity, pi_gate_process_fidelity
STALE (re-measure or re-derive): drag_beta, T1_us, T2_us, T2star_us
=================================================
\end{Verbatim}
\vspace{-0.3em}
\caption{The calibration notebook injected on every turn, captured verbatim from a hard-tier two-qubit edge sub-episode (evidence summaries abridged; \code{r^2} shown for $r^2$). A separate representative example, not the same episode as the plots or traces above. Each committed parameter carries its value and a one-line evidence record (setting tool, fit $r^2$); the header tracks modeled wall time, and \code{[age:N min]} tags flag aging values. The coherence and DRAG values inherited from the single-qubit stage (\code{prior_calibration}) are marked \code{[STALE]} after an upstream re-measurement, flagging them for re-derivation; and the iSWAP chevron's failed fit ($r^2{=}-0.07$, \code{recommended=null}) is why its duration was set from the physical prior $1/(2g)$ rather than the fit. This block, not the scrolling transcript, is the state the prompt directs the agent to trust.}
\label{fig:app-notebook}
\end{figure}

\end{document}